\documentclass{ecai} 

\usepackage{latexsym}
\usepackage{amssymb}
\usepackage{amsmath}
\usepackage{amsthm}
\usepackage{booktabs}
\usepackage{enumitem}
\usepackage{graphicx,subfig}
\usepackage{color}
\usepackage{adjustbox,bbm}

\usepackage{algorithmic}
\usepackage{algorithm}

\usepackage{amsmath,amsfonts,amssymb,amsthm}
\usepackage{multirow}

\newtheorem{Gam}{Game}
\newtheorem{Def}{Definition}
\newtheorem{The}{Theorem}
\newtheorem{Pro}{Proposition}
\newtheorem{Cor}{Corollary}
\newtheorem{Assump}{Assumption}

\newtheorem{Lem}{Lemma}
\newtheorem{Obs}{Observation}
\newtheorem{question}{Question}

\newcommand{\BibTeX}{B\kern-.05em{\sc i\kern-.025em b}\kern-.08em\TeX}

\begin{document}


\begin{frontmatter}


\paperid{123} 


\title{Technical Report: Coopetition in Heterogeneous Cross-Silo Federated Learning}

\author{Chao Huang, Justin Dachille, and Xin Liu
}



\begin{abstract}
In cross-silo federated learning (FL), companies collaboratively train a shared global model without sharing heterogeneous data. Prior related work focused on algorithm development to tackle data heterogeneity. However, the dual problem of \textit{coopetition}, i.e., FL collaboration and market competition, remains under-explored. This paper studies the FL coopetition using a dynamic two-period game model. In period 1, an incumbent company trains a local model and provides model-based services at a chosen price to users. In period 2, an entrant company enters, and both companies decide whether to engage in FL collaboration and then compete in selling model-based services at different prices to users. Analyzing the two-period game is challenging due to data heterogeneity, and that the incumbent's period one pricing has a temporal impact on coopetition in period 2, resulting in a non-concave problem.  To address this issue, we decompose the problem into several concave sub-problems and develop an algorithm that achieves a global optimum. Numerical results on three public datasets show two interesting insights. First, FL training brings model performance gain as well as competition loss, and collaboration occurs only when the performance gain outweighs the loss. Second, data heterogeneity can incentivize the incumbent to limit market penetration in period 1 and promote price competition in period 2. 
\end{abstract}

\end{frontmatter}

\section{Introduction}\label{intro}
Cross-silo federated learning (FL) is a distributed machine learning paradigm where multiple companies or organizations train a shared model collaboratively without directly exchanging local data \cite{kairouz2021advances}. Typically, the process involves each participant training a local model on their dataset and then sharing model updates with a coordinating server that aggregates these updates to improve a global model. This local training and aggregation iteration continues until the global model converges \cite{huang2021personalized}. Potential application scenarios for cross-silo FL are abundant \cite{huang2023promoting}. For example, in healthcare, hospitals can collaborate on medical research (e.g., disease diagnosis) without sharing sensitive patient data. In finance, different banks can use FL for improved fraud detection without exposing customer data. In smart manufacturing, companies can enhance predictive maintenance without revealing proprietary operational data.

While a significant volume of recent research has focused on improving the model performance in cross-silo FL (e.g., \cite{zhang2020batchcrypt,liu2022privacy,marfoq2020throughput}), the critical aspect of \textit{market competition} remains less explored. In practice, companies can utilize the shared global model developed through FL to offer model-based services, hence competing for the same pool of potential users \cite{wu2022mars}. This competition can manifest across various industries. In healthcare, hospitals may compete to attract patients by offering diagnostic services or personalized treatment plans. In finance, banks might leverage an enhanced fraud detection system to attract customers by offering greater security and reduced risk of fraud. In smart manufacturing, companies could use the shared global model to optimize operations and attract clients/partnerships via more efficient production services. The dual focus of FL collaboration and market competition is termed as \textbf{FL coopetition}, which is the focus of this paper. 

\begin{figure}[t]
	\centering
        \includegraphics[width=3.5in]{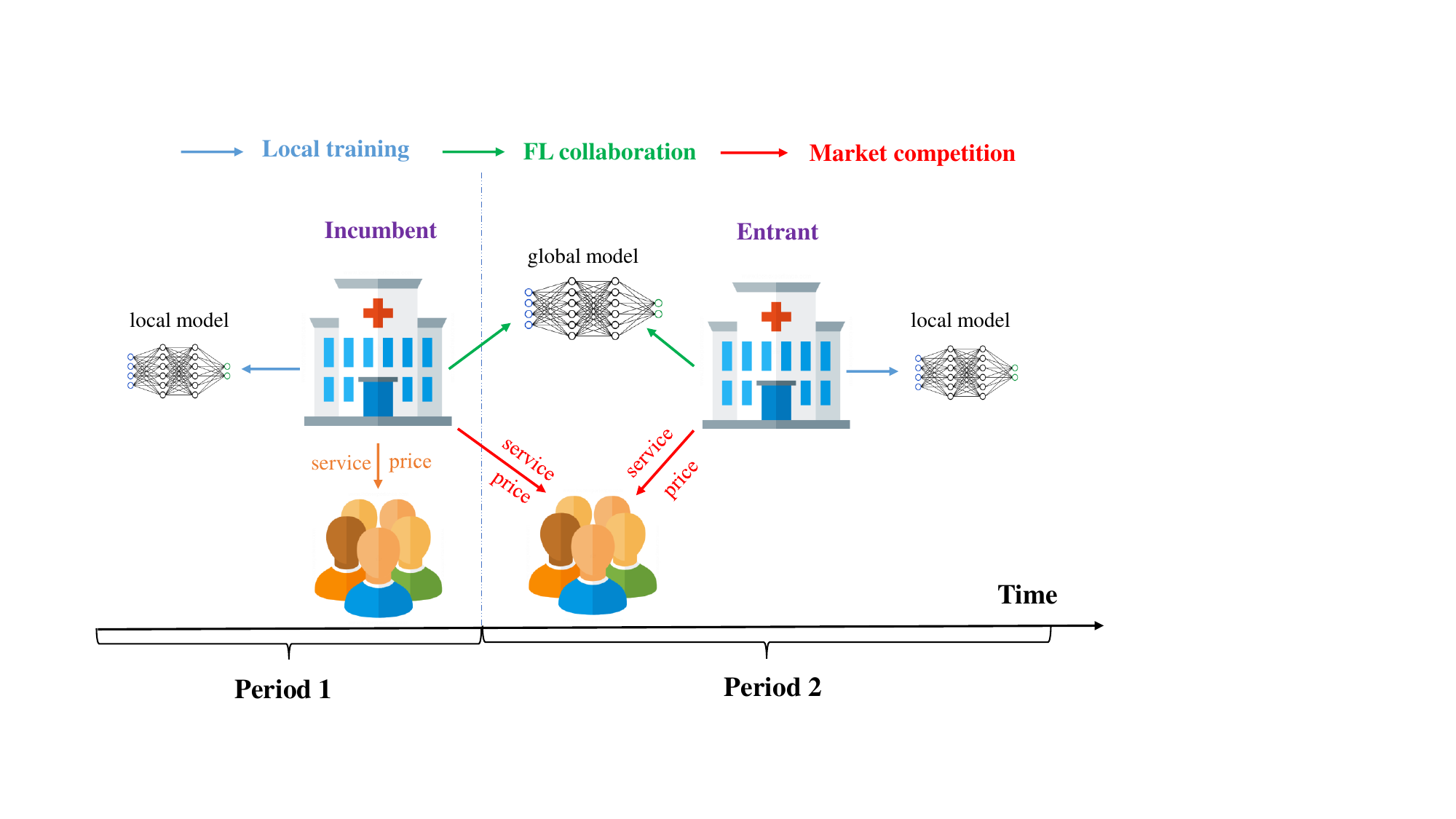}
	\caption{The two-period system model.}
	\label{fig-system-model}
 \vspace{5mm}
\end{figure}

While a few recent studies looked at FL coopetition \cite{huang2024duopoly,tan2024fedcompetitors,tsoy2024strategic,wu2022mars}, they focused on static competition and overlooked the important aspect of dynamic market entry. 
Market entrant usually serves as a catalyst for technology innovation (e.g., model and algorithm development in cross-silo FL) and helps provide insights into the interactions between incumbents and entrants \cite{duffy2005learning}. This also impacts the socioeconomic landscape by promoting coopetition and increasing the quality of services (via FL) available to users in the market.

In this paper, we study FL coopetition using a dynamic market entry model that spans two time periods involving three entities: users, an incumbent company, and an entrant company (see Fig.~\ref{fig-system-model}):
\begin{itemize}
\item Period 1: Only an incumbent company exists in the market, providing users with the model-based service. For example, consider a healthcare provider that has developed a new AI-based tool for early diagnosis of a specific disease and is the only service provider in the market. The company trains a local model on its dataset and offers services to patients at a chosen price.
\item Period 2: An entrant company enters the market and coexists with the incumbent in providing user services. The incumbent and the entrant need to decide strategically on whether to engage in FL collaboration. If they collaborate, they will use a jointly developed global model to enhance and offer their diagnostic services. If they opt against collaboration, each will continue using individually trained models. In addition, since users can now choose between different service providers, there is a price competition between the incumbent and the entrant.
\end{itemize}
Given this dynamic model of FL coopetition, our first question is:
\begin{question}\label{question-1}
Will the incumbent collaborate with the entrant via FL in the presence of competition? 
\end{question}
To answer Question \ref{question-1}, we model the FL collaboration and market competition as two (intertwined) games between the incumbent and the entrant. Each company aims to maximize its profit (from selling model-based services to users). We are interested in solving the game-theoretical equilibrium, which is highly non-trivial due to two reasons. First,  companies in practice are heterogeneous regarding their data distributions and quantities \cite{qin2023fedapen}. Second,  users may exhibit heterogeneous preferences towards companies' services (even if the services have similar qualities). This can happen when some users find that a particular company's service matches more closely with their individual characteristics, e.g., demographic profiles and personal attributes \cite{osborne1987equilibrium}. It is challenging to analyze (or even model) the game. To address this issue, we resort to the renowned Hotelling model in economics \cite{hotbllino1929stability} and generalize it to model various types of heterogeneity. We characterized the equilibrium with arbitrary data distributions/quantities of companies and under minor assumptions about users' preferences.

Among aforementioned heterogeneity types, data heterogeneity receives the most research attention from the FL community and continues to be the major bottleneck \cite{ye2023heterogeneous}. There are many excellent studies on algorithm development to mitigate the client drift issue caused by data heterogeneity, e.g., \cite{li2021model,li2020federated,son2024feduv,li2021ditto}. We will show that our model and analysis are orthogonal to any FL algorithms.  Since we focus on FL coopetition, our second key question is:
\begin{question}\label{question-2}
How does data heterogeneity affect FL collaboration and market competition?	
\end{question}
To answer Question \ref{question-2}, we solve the equilibrium of the two-period game and conduct numerical experiments on three public datasets under various levels of data heterogeneity. Notice that the decision-making of the incumbent's pricing in period 1 presents challenges due to its temporal impacts. That is, it not only influences user decisions in period 1, but also affects the decisions of both companies regarding FL collaboration (with heterogeneous data) and price competition, as well as user decisions in period 2. We will show that this interdependence makes the equilibrium analysis a challenging non-concave problem.


Our key contributions are summarized as follows:
\begin{itemize}
\item To our best knowledge, this is the first work to study FL coopetition from a temporal market entry perspective. We formulate a generic two-period model that accommodates heterogeneity including companies' data distributions and quantities as well as users' service preferences. Our model is also orthogonal to any FL algorithms.
\item We provide game-theoretical solutions to the two-period model. The analysis involves solving a non-concave problem. 
To tackle this challenge, we decompose the  problem into several manageable concave sub-problems and further develop an algorithm that achieves the global optimum. 
\item We conduct numerical experiments on three public datasets and show two interesting results. First, even under highly heterogeneous data,  FL training improves model performance (compared to local learning). However, companies may avoid collaborating in the presence of competition, as FL collaboration also benefits the competitor. Second,  data heterogeneity can incentivize the incumbent to limit market penetration in period 1 and promote price competition in period 2. 
\end{itemize} 

\subsection{Related Work}
  \textbf{Heterogeneous FL}. Data heterogeneity, also commonly referred to as non-IID data,\footnote{Non-IID means not identically and independently distributed.} is known to cause significant performance loss in FL due to the client drift issue. Many studies have focused on algorithm development to tackle client drift. One commonly used technique is regularization by adding proximal terms to restrain local updates with respect to the global model, e.g., FedProx \cite{li2020federated}, SCAFFOLD \cite{karimireddy2020scaffold}, MOON \cite{li2021model}, FedUV \cite{son2024feduv}. Other works have approached the problem using personalization (e.g., \cite{oh2022fedbabu,pillutla2022federated}) and local learning generality (e.g., \cite{mendieta2022local}). 

This paper focuses on how data heterogeneity affects FL coopetition.  Our model and analysis are orthogonal to any FL algorithms mentioned above and beyond. 

\textbf{Incentives for cross-silo FL}. Our work is related to game theoretical and, in particular, incentive studies for cross-silo FL, e.g., \cite{tang2021incentive,zhang2022enabling,huang2022incentivizing,li2023varf,huang2023promoting}. For example, Tang and Wong \cite{tang2021incentive} proposed an auction-like mechanism to encourage organizations' FL training participation. Zhang et al. \cite{zhang2022enabling} studied how organizations participate in long-term collaboration. However, these studies overlooked the important aspect of market competition. 

Only until very recently, a few papers have looked at market competition in cross-silo FL \cite{huang2024duopoly,tan2024fedcompetitors, wu2022mars,tsoy2024strategic}. Huang et al. \cite{huang2024duopoly} studied an oligopoly market. Wu and Yu \cite{wu2022mars} focused on a fully competitive market. 
Tan \cite{tan2024fedcompetitors} developed an algorithm to find stable collaboration structures among companies. Tsoy and Konstantinov \cite{tsoy2024strategic} studied both the price and quantity competitions in cross-silo FL. However, these studies did not study how  data heterogeneity affects competition. More importantly, they looked at a static competition model while we study a dynamic model where the the incumbent and entrant have temporal interactions.  The dynamic model renders the problem a challenging non-concave one. 

\textbf{Market competition in economics}. Market competition is an extensively studied topic in economics, e.g., see \cite{tirole1988theory,stigler1964theory} for comprehensive discussions on theories of competition. Most pertaining to our work is the renowned Hotelling model (e.g.,  \cite{osborne1987equilibrium,570930af-c28a-3bc3-be3a-5052d0478611}), where companies at different spatial locations compete for users. Our model and analysis differ from the conventional Hotelling due to the unique features of data heterogeneity and collaborative training in FL. That is, prior studies considered that each company can independently decide their service qualities to attract users. In cross-silo FL, however, the qualities of companies' model-based services are dependent on their heterogeneous data distributions and the FL collaboration strategies. This makes our analysis more challenging than and conclusions different from prior literature.

\textit{In summary, to our best knowledge, this paper is the first to study FL coopetition using a dynamic market entry model and also considering data heterogeneity.}

The rest of the paper is organized as follows. Sec.~\ref{model} introduces the system model, and Sec.~\ref{solution} provides both analytical and algorithmic solutions. Sec.~\ref{numerical} presents the numerical results. We discuss the extensions of our model in Sec.~\ref{Discussions} and conclude in Sec.~\ref{Conclusions}.

\section{Model}\label{model}
Sec.~\ref{sec: company-users} introduces the companies and users.  Sec.~\ref{sec:two-period-game} models their two-period game-theoretical interactions.

\subsection{Companies and Users}\label{sec: company-users}

\textbf{Companies}. We consider a two-period interaction with time index $t\in \mathcal{T}=\{1,2\}$. In period $t=1$, there is an \underline{i}ncumbent company $I$ who holds a private dataset $\mathcal{D}_I$ with size $D_I=|\mathcal{D}_{I}|$. It trains a machine learning model using $\mathcal{D}_I$ and provides model-based services (e.g., disease diagnosis) to users at a price $p_{I,1}$. In period $t=2$, an \underline{e}ntrant company $E$ who has a local private dataset $\mathcal{D}_E$ with size $D_E=|\mathcal{D}_E|$ enters the market.\footnote{See Sec. \ref{Discussions} for discussions on how to extend our model and analysis 1) with more than two companies, and 2) across $t\ge 3$ periods. 
} Since both companies $I$ and $E$ are in the market, they can choose to collaborate via FL to improve the performance of their models and the quality of model-based services. We use $r_i\in \{1,0\}, i \in \{I,E\}$ to denote the collaboration decision, where $r_i=1$ means FL collaboration and $r_i=0$ means no collaboration. More specifically, 
\begin{itemize}
\item If $r_I \cdot r_E=1$, both companies train a shared global model using $\mathcal{D}_I$ and $\mathcal{D}_E$ till convergence. They will use the converged model to generate model-based services.
\item If $r_I \cdot r_E=0$, each company uses its own data to train a local model, based on which services are generated.
\end{itemize}
We use $q_{i,t}$ to denote company $i$'s quality of service in period $t$. In particular, $q_{i,2}$ is a function of $\boldsymbol{r}=\{r_I, r_E\}$ that depends on both companies' decisions on FL collaboration (and data heterogeneity), which we refer to as $q_{i,2}(\boldsymbol{r})$ hereafter.

Besides the potential FL collaboration, both companies compete for the same users at prices $p_{i, 2}$. For simplicity, let $\boldsymbol{p}=\{p_{I,1}, p_{I,2}, p_{E,2}\}.$ Next, we introduce how users respond to companies' model-based services and prices.   

\textbf{Users}. The market consists of a large number of users. To model this, we consider a continuum of users and normalize the total mass to be one. Each user decides whether to buy the service, and if so, from which company. Similar to \cite{ma2018dynamic}, we assume that each user buys at most once, i.e., if a user buys from company $I$ in period 1, it will not buy from company $I$ or $E$ in period 2 due to them offering substitutable services. Let $d_{n,t} \in \{\emptyset,i\}$ denote user $n$'s purchasing decision in period $t$, where $d_{n,t}=\emptyset$ means no purchasing and $d_{n,t}=i$ means purchasing from company $i$. 

A user's payoff consists of three parts discussed below.
\begin{itemize}
\item First, a user obtains a utility from enjoying the model-based service, where the utility increases in the quality of service $q_{i,t}$. This paper considers a linear function (e.g., \cite{ma2018dynamic}). It is easy to extend the model to an arbitrary non-decreasing function (similar to \cite{huang2024duopoly}).  
\item Second, users have  \textit{heterogeneous preferences} towards companies' services. This can arise due to the traveling cost or brand loyalty over a particular company \cite{osborne1987equilibrium}. In the context of cross-silo FL, some users may find that a company's service matches more closely with their individual characteristics (e.g., demographic profiles and personal attributes).  To model users' heterogeneous preferences, we use the renowned Hotelling model \cite{hotbllino1929stability,tirole1988theory}, in which users are located on a line $[0,1]$ and companies $I$ and $E$ are at the ending points $\phi=0$ and $\phi=1$, respectively. The users' locations $\phi_n$ follow a distribution with a PDF $h(\phi)$ and a CDF $H(\phi)$, which are known to both companies due to market research \cite{ma2018dynamic,huang2024duopoly}. Here,  we model a user's \textit{preference misalignment} as the distance to the company where the service is purchased.\footnote{The Hotelling model and its variants (e.g., \cite{salop1979monopolistic}) have been widely used to model user preferences in the economics literature.} One can understand such preference misalignment as the dissimilarity between users' characteristics and the companies' data distributions. 
\item Third, a user needs to pay a price for the service.
\end{itemize}
    Next, we define a user $n$'s payoff as follows:
\begin{equation}
\begin{aligned}
	&u_{n,t} (d_{n,t}; \boldsymbol{r}, \boldsymbol{p})\\ 
 &= \begin{cases} 0, & \text{if } d_{n,t} = \emptyset,\\ w_q q_{I, t}(\boldsymbol{r}) - \hspace{10mm}w_\phi \phi_n \hspace{9mm}- w_p p_{I,t}, & \text{if } d_{n,t} = I,  \\
    \underbrace{w_q q_{E, t}(\boldsymbol{r})}_{\rm utility} - \underbrace{w_\phi (1-\phi_n)}_{\rm preference~misalignment} - \underbrace{w_p p_{E,t}}_{\rm price}, & \text{if } d_{n,t} = E,  \\
      \end{cases}
      \end{aligned}
\end{equation}
where $w_q, w_\phi, w_p$ are positive constants. Without loss of generality, we normalize $w_{\phi}$ to $1$. Note that in period 1, only company $I$ is in the market, so a user can only choose to buy from $I$ or not buy.

\subsection{The Two-Period Game}\label{sec:two-period-game}
\begin{figure}[t]
	\centering
        \includegraphics[width=3in]{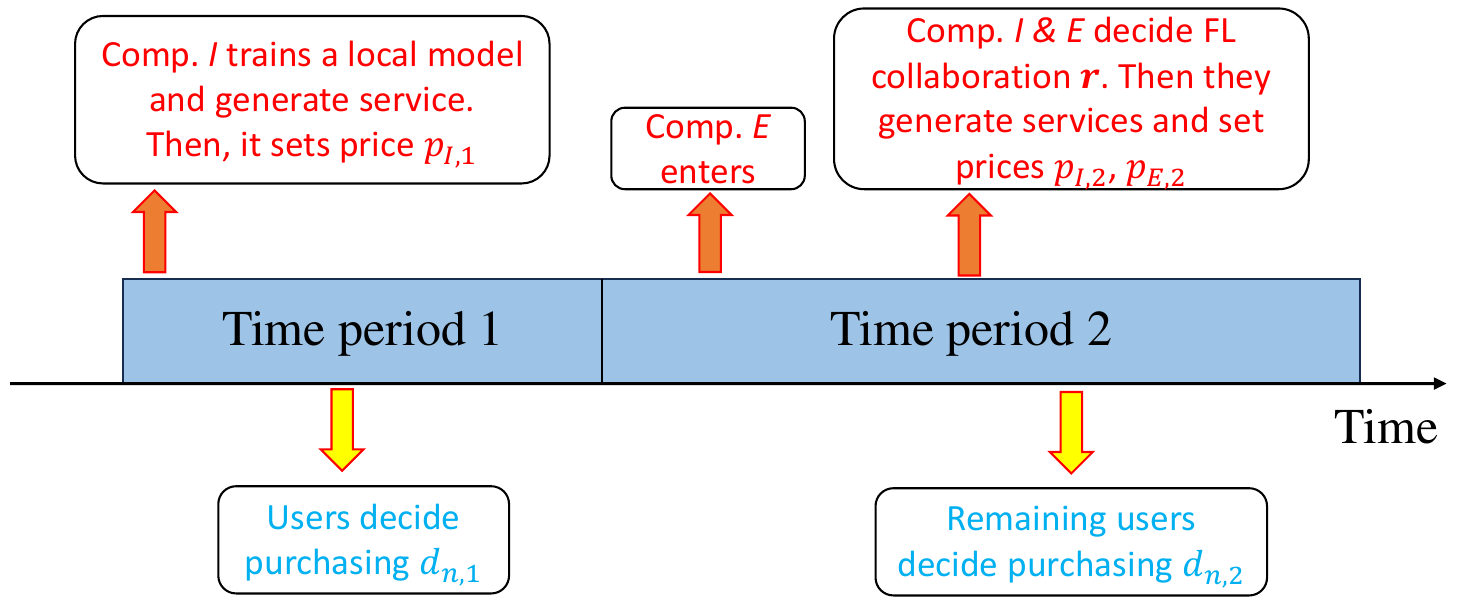}
	\caption{The two-period game.}
	\label{fig-two-period-game}
 \vspace{5mm}
\end{figure}

We model the interactions among the two companies and users as a two-period game (see Fig.~\ref{fig-two-period-game}). At the beginning of period 1, company $I$ provides service using locally trained model with quality $q_{I,1}$ at price $p_{I,1}$. Then, users make purchasing decisions. In period 2, company $E$ enters. Both companies decide whether to collaborate via FL $\boldsymbol{r}$, based on which they generate model-based services with quality $q_{i,2}(\boldsymbol{r})$. Then, they initiate price competition by selling services at prices $p_{i,2}$.  

\textbf{Users' decision problem}. We consider that users are \textit{myopic} \cite{kremer2017dynamic,ma2018dynamic}. That is, in period 1, if a user finds purchasing from company $I$ is better off than not purchasing, it will purchase without anticipating the potential entry of company $E$ and its service in period 2. Hence, each user's decision problem is as follows. 
\begin{align}
	\textbf{P1}: \quad\quad \max_{d_{n,t}} \;\;\; u_{n,t} (d_{n,t}; \boldsymbol{r}, \boldsymbol{p}), \quad \quad  \forall t.
\end{align}

\textbf{Companies' decision problem}. Both companies aim to maximize their own profit. For company $E$, its goal is to maximize the profit obtained in period 2, i.e., 
\begin{equation}\label{E-profit-function}
W_E(\boldsymbol{r}, \boldsymbol{p}) = \int_{0}^{1} (p_{E,2}-c_{E}) \cdot \mathbbm{1}_{d_{n,2}=E}(\boldsymbol{r}, \boldsymbol{p}) \cdot h(\phi) d\phi,
\end{equation}
where  \( \mathbbm{1} \) is an indicator function, meaning \( \mathbbm{1}_{d_{n,2}=E} = 1 \) if and only if \( d_{n,2}=E \). Here, $c_{E}$ is the marginal service cost.\footnote{Similar to \cite{huang2023promoting}, this paper normalizes the FL collaboration cost, in terms of computation and communication, to be zero. Our analysis and conclusions will not change even if we consider a non-zero cost.} Hence, company $E$'s problem is as follows.
\begin{equation}
\textbf{P2}: \quad\quad \max_{r_E, p_{E,2}} \;\;\; W_E(\boldsymbol{r}, \boldsymbol{p}).
\end{equation}
Unlike users who are myopic, company $I$ is assumed to be forward-looking \cite{spann2015skimming}.\footnote{This is reasonable due to companies usually having more resources and information than individual users when making decisions.} That is, it will strategically set the price in period 1, anticipating the reactions of users and company $E$ in period 2. Here, we consider that company $I$ knows the information about $q_{i,2}(\boldsymbol{r}), \forall \boldsymbol{r}$, which can be acquired via market research \cite{ma2018dynamic,huang2024duopoly}.   In particular, company $I$'s aims to maximize its two-period profits:
\begin{equation}\label{I-profit-period-1}
W_{I,1}(p_{I,1}) = \int_{0}^{1} (p_{I,1}-c_{I}) \cdot \mathbbm{1}_{d_{n,1}=I}(p_{I,1}) \cdot h(\phi) d\phi,
\end{equation}
\begin{equation}\label{I-profit-function-2}
W_{I,2}(\boldsymbol{r}, \boldsymbol{p}) = \int_{0}^{1} (p_{I,2}-c_{I}) \cdot\mathbbm{1}_{d_{n,2}=I}(\boldsymbol{r}, \boldsymbol{p}) \cdot h(\phi) d\phi,
\end{equation}
where $c_{I}$ is the marginal service cost. Hence, we formulate company $I$'s problem below.
\begin{equation}\label{I-total-profit}
\textbf{P3}: \quad\quad \max_{r_I, p_{I,1}, p_{I,2}} \;\;\;W_{I}(\boldsymbol{r}, \boldsymbol{p})=W_{I,1}(p_{I,1})+W_{I,2}(\boldsymbol{r}, \boldsymbol{p}).
\end{equation}		
Next, we solve the two-period game.

\section{Solving the Two-Period Game}\label{solution}
We first solve the user decision in Sec.~\ref{solution-user}. Then, we solve the companies' period 2 decisions in Sec.~\ref{sec-company-period-2}. Sec.~\ref{sec-company-period-1} presents an algorithm to solve company $I$'s pricing in period 1. The proofs for all technical results are left to Appendices at the end of this paper. 
\subsection{Users' Optimal Purchasing in Two Periods}\label{solution-user}
Since users are myopic, we can solve users' optimal purchasing in the two periods separately. 

\begin{Lem}\label{user-period-1}
	Each user $n$'s optimal purchasing in period 1 is 
	\begin{equation}\label{purchasing-period-1}
		\begin{aligned}
			d^{*}_{n,1}(p_{I,1}) =
             \begin{cases}
			I,  \;{\rm if}  \; \phi_n\le w_q q_{I,1}-w_pp_{I,1}, \\
			\emptyset,  \;{\rm else}.
		\end{cases}
	\end{aligned}
	\end{equation}
	\end{Lem}
Lemma \ref{user-period-1} shows that a user is more likely to purchase from company $I$ if the model-based service has a better quality $q_{I,1}$, or the price $p_{I,1}$ is lower, or the preference misalignment $\phi_n$ is smaller.

\begin{figure*}[t]
\begin{equation}\label{user-period-2}
\begin{aligned}
        &d^*_{n,2}(\boldsymbol{r}, \boldsymbol{p}) \\
        &\hspace{-3mm}=\begin{cases}
     I,  &\hspace{-2mm}{\rm if} \; d^*_{n,1}(p_{I,1})=\emptyset,  \phi_n\in \left[0,\max \left(\min \left(w_q q_{I,2}(\boldsymbol{r})-w_pp_{I,2},\frac{1+w_q\left(q_{I,2}(\boldsymbol{r})-q_{E,2}(\boldsymbol{r})\right)-w_p(p_{I,2}-p_{E,2})}{2},1\right),0 \right)\right],\\
     E, &\hspace{-2mm}{\rm if}\; d^*_{n,1}(p_{I,1})=\emptyset, \phi_n \in \left[\min \left(\max \left(1-w_q q_{E,2}(\boldsymbol{r})+w_pp_{E,2},\frac{1+w_q\left(q_{I,2}(\boldsymbol{r})-q_{E,2}(\boldsymbol{r}\right)-w_p(p_{I,2}-p_{E,2})}{2},0\right),1\right) ,1\right], \\
     \emptyset, \ & \hspace{-2mm}{\rm else}.
      \end{cases}
      \end{aligned}
\end{equation}
\end{figure*}

\begin{Lem}\label{Lem-user-2}
Assume $w_qq_{I,2}(\boldsymbol{r})-w_pp_{I,2}\ge w_qq_{E,2}(\boldsymbol{r})-w_pp_{E,2}$.
Each user $n$'s optimal purchasing in period 2 is given in (\ref{user-period-2}). 
\end{Lem}
The assumption in Lemma \ref{Lem-user-2} is without loss of generality. One can similarly analyze the case where $w_qq_{I,2}(\boldsymbol{r})-w_pp_{I,2}< w_qq_{E,2}(\boldsymbol{r})-w_pp_{E,2}$. Importantly, Lemma \ref{Lem-user-2} shows that users' purchasing in period 2 depends on companies' FL collaboration $\boldsymbol{r}$ and their prices $\boldsymbol{p}$, including company $I$'s period 1 price $p_{I,1}$. As will be shown, such temporal dependence will render company $I$'s decision problem challenging.

\subsection{Companies' Decisions in Period 2}\label{sec-company-period-2}
In period 2, given $p_{I,1}$ in period 1, both companies decide the FL collaboration $\boldsymbol{r}$ and prices $\{p_{I,2}, p_{E,2}\}$ to maximize profits. We solve the pricing in Sec.~\ref{sec-pricing-period-2} and FL collaboration in Sec.~\ref{sec-FL-period-2}.

\subsubsection{Companies' Pricing in Period 2}\label{sec-pricing-period-2}
We first solve companies' optimal pricing in period 2, given $p_{I,1}$ and $\boldsymbol{r}$. 
Since both companies' prices will affect users' purchasing decisions and hence the company profits, the companies are playing a price competition game, which we model as follows. 
\begin{Gam}{(Price Competition in Period 2)}\label{price-game}
 The price competition game in period 2 is defined as a tuple $\langle\{I,E\}, \mathcal{P}=\prod p_{i,2}, \mathcal{W}=\prod W_{i,2}\rangle$, where each company $i$ in $\{I,E\}$ decides the pricing $p_{i,2}$ to maximize its own profit $W_{i,2}$ in (\ref{E-profit-function}) and (\ref{I-profit-function-2}).
	\end{Gam}
We aim to find the Nash equilibrium  (NE) of Game \ref{price-game}.
\begin{Def}
	A  profile $\boldsymbol{p}_2^*(\boldsymbol{r},p_{I,1})=(\boldsymbol{p}_{i,2}^*(\boldsymbol{r},p_{I,1}),\boldsymbol{p}_{j,2}^*(\boldsymbol{r},p_{I,1}))$ is an NE of Game \ref{price-game} if for all $i\in \{I,E\}$, $p_{i,2}'(\boldsymbol{r},p_{I,1}) \neq p_{i,2}^*(\boldsymbol{r},p_{I,1})$,
	\begin{equation}
		W_{i,2}(p_{i,2}^*(\boldsymbol{r},p_{I,1}), p_{j,2}^*(\boldsymbol{r},p_{I,1}))\ge W_{i,2}(p_{i,2}'(\boldsymbol{r},p_{I,1}), p_{j,2}^*(\boldsymbol{r},p_{I,1})), 
	\end{equation}
where $j\neq i$ and $j\in \{I,E\}$.
\end{Def}
NE is considered as a stable strategy profile as no company can achieve a higher profit via unilaterally changing its pricing. Notice that an arbitrary choice of $H(\cdot)$ can easily render the equilibrium analysis intractable. To facilitate theoretical analysis, we make some minor assumptions on $H(\cdot)$ (and  $h(\cdot)$). 
\begin{Assump}\label{minor-assumption}
Users' preference distribution satisfies:	(i) $h(\theta)>0$ and is continuous. (ii) $h(\theta)/[1-H(\theta)]$ is increasing in $\theta$.
\end{Assump}

Assumption \ref{minor-assumption} holds for many widely used distributions, e.g., Gaussian, uniform, and gamma distributions. Now, we discuss the equilibrium existence below. 

\begin{Pro}\label{price-game-NE-existence}
Under Assumption \ref{minor-assumption}, Game \ref{price-game}'s NE exists. 
\end{Pro}
Proposition \ref{price-game-NE-existence} can be proved by showing that the price competition game is a concave game, and in particular, each organization's profit is quasi-concave in its price. Refer to the supplementary material where we have developed a best response algorithm to compute the price equilibrium. 

Next, we provide a result that establishes the relationship between the optimal pricing and companies' quality of service. 
\begin{Cor}\label{cor-price-quality-relation}
Each company $i$'s optimal pricing in period 2 is a non-decreasing function in $q_{i,2}(\boldsymbol{r})$ but a non-increasing function in $q_{j,2}(\boldsymbol{r})$, where $i,j\in \{I,E\}, i\neq j$.
\end{Cor}
Corollary \ref{cor-price-quality-relation} can be proved via contradiction. It means that each company will set a higher price if it has a better quality of service, while it will lower the price if its competitor has a better quality. 

\subsubsection{Companies' FL Collaboration in Period 2}\label{sec-FL-period-2}
Now, we solve companies' optimal FL collaboration in period 2, given $p_{I,1}$ in period 1. The companies decide  collaboration strategy $\boldsymbol{r}$ for FL, anticipating the price equilibrium of Game \ref{price-game}. 
We model the two companies' interactions as a collaboration game.
\begin{Gam}{(FL Collaboration Game in Period 2)}\label{collaboration-game}
	The FL collaboration game is a tuple $\langle\{I,E\}, \mathcal{R}=\prod r_i, \mathcal{W}=\prod W_{i,2}\rangle$, where each company $i$  decides its FL collaboration $r_i$ to maximize $W_{i,2}(\boldsymbol{r}, \boldsymbol{p}_{2}^*(\boldsymbol{r}, p_{I,1}))$, where $\boldsymbol{p}_{2}^*(\boldsymbol{r}, p_{I,1})$ is the NE of Game \ref{price-game}.
\end{Gam}
We aim to solve Game \ref{collaboration-game}'s NE defined below.
\begin{Def}
	A strategy profile $\boldsymbol{r}^*=(r_i^*,r_{j}^*)$ is an NE of Game \ref{collaboration-game} if 
	 $\forall i\in \{I,E\}$, $\forall r_i'\neq r_i^*$,
	\begin{equation}
		W_{i,2}( \boldsymbol{r}^*, \boldsymbol{p}_2^*(\boldsymbol{r}^*, p_{I,1})) \hspace{-1mm}\ge \hspace{-1mm}W_{i,2}( (r_i', r_j^*), \boldsymbol{p}_2^*((r_i', r_j^*), p_{I,1}), 
	\end{equation}
	where $j\neq i$ and $j\in \{I,E\}$. 
\end{Def}

Next, we characterize the NE in Proposition\ref{collaboration-NE}.

\begin{Pro}\label{collaboration-NE}
 The profile $(1, 1)$ is the NE of Game \ref{collaboration-game} if and only if $\forall i\in \left\{I,E\right\}$, $W_{i,2}( (1,1), \boldsymbol{p}_2^*((1,1), p_{I,1})) \ge W_{i,2}((0, 0), \boldsymbol{p}_2^*((0,0), p_{I,1})).$
Otherwise, $(0,0)$ is the NE.
\end{Pro}

Proposition \ref{collaboration-game} means that the two companies will collaborate in FL training if and only if both achieves a higher profit than no collaboration. 
Next, we characterize a somewhat counter-intuitive result. 
\begin{The}\label{NE-GAME-2}
Under Assumption \ref{minor-assumption}, there exists a $p_{I,1}$ such that $(0,0)$ is the NE of Game \ref{collaboration-game} even if $ q_{i,2}((1,1))>q_{i,2}((0,0)), \forall i$. 
\end{The}
Theorem \ref{NE-GAME-2} shows that even if FL collaboration improves both companies' quality of service, it may not be the NE. This is because FL brings model performance gain and \textit{competition loss} at the same time. That is, the shared FL model also improves the competitor's quality of service.  According to Corollary \ref{cor-price-quality-relation}, a company needs to set a lower price to attract users, which can lead to a lower profit. We will show in Sec.~\ref{numerical} that FL is the NE only when the model performance gain is significant and outweighs the competition loss.

\subsection{Company $I$'s Pricing in Period 1}\label{sec-company-period-1}
 We solve company $I$'s optimal pricing in period 1. In particular, company $I$ decides $p_{I,1}$ to maximize its total profit in (\ref{I-total-profit}), anticipating the equilibria of the price competition game and FL collaboration game.  

Note that solving this problem is challenging due to a few reasons. First, company $I$'s period-1-price $p_{I,1}$ affects users' decisions in both periods 1 and 2 (see (\ref{purchasing-period-1})-(\ref{user-period-2})). This in turn will affect companies' price competition and FL collaboration in period 2, leading to a highly coupled analysis. Second, even if we established equilibrium existence in Propositions \ref{price-game-NE-existence}-\ref{collaboration-NE}, we still lack closed-form equilibrium solutions, making the characterization of the profit function challenging. Third, it is easy to showcase that company $I$'s total profit is non-concave  in $p_{I,1}$ if we use a Gaussian distribution for $\phi_n$.


To address this challenge, we develop an algorithm to compute company $I$'s optimal pricing in period 1 in Algorithm \ref{alg:B}. For ease of presentation, we use $\text{NE}_p(p_{I,1},l,h)$ and $\text{NE}_{r}(p_{I,1},l,h)$ to denote the equilibrium  calculated on price $p_{I,1}\in [l,h]$ of Game \ref{price-game} and Game \ref{collaboration-game}, respectively. The calculation of $\text{NE}_p$ is based on the a best response algorithm (details in appendix). The calculation of $\text{NE}_r$ is based on Proposition \ref{collaboration-NE}. Note that even if the problem is non-concave, the proposed algorithm can return the global optimal solution.
\begin{The}
Under Assumption \ref{minor-assumption}, Algorithm \ref{alg:B} converges to company $I$'s optimal pricing in period 1.
\end{The}
The key rationale is that even if the problem is non-concave, we can decompose the problem into several concave sub-problems (i.e., sub-problem $a$ in lines 2-6, sub-problem $b$ in lines 7-11, and sub-problem $c$ in lines 12-16). Then it suffices to locate the optimal solutions for each sub-problem and then compare those to achieve the global optimum.

\begin{table*}
\centering
\caption{Impact of data heterogeneity $\beta$ on model accuracy in $\%$. }\label{table-beta}
\begin{adjustbox}{max width=\textwidth}

\begin{tabular}{|c|c|c|c|c|c|} \hline
\multicolumn{6}{|c|}{CIFAR10} \\ \hline
& $\beta=\infty$ & $\beta=1.0$ & $\beta=0.5$& $\beta=0.1$ & $\beta=0.01$ \\ \hline
$I$ local & 60.81 ± 4.32 & 55.34 ± 1.82 & 53.42 ± 5.37 & 39.23 ± 6.84 & 43.27 ± 4.03 \\ \hline
$E$ local & 57.56 ± 3.75 & 54.50 ± 3.58 & 49.30 ± 7.41 & 48.17 ± 5.93 & 33.51 ± 4.95 \\ \hline
FedAvg & 69.81 ± 2.26 & 69.01 ± 2.22 & 66.96 ± 2.77 & 57.40 ± 9.02 & 51.89 ± 6.49 \\ \hline
\end{tabular}

\vspace{1em} 

\begin{tabular}{|c|c|c|c|c|c|} \hline
\multicolumn{6}{|c|}{CIFAR100} \\ \hline
& $\beta=\infty$ & $\beta=1.0$ & $\beta=0.5$& $\beta=0.1$ & $\beta=0.01$ \\ \hline
 $I$ local& 19.00 ± 4.28 & 18.50 ± 2.79 & 17.37 ± 2.11 & 17.17 ± 3.06 & 16.94 ± 1.89 \\ \hline
 $E$ local& 17.93 ± 2.78 & 17.46 ± 2.39 & 17.12 ± 2.66 & 18.09 ± 4.49 & 14.26 ± 1.79 \\ \hline
FedAvg & 32.80 ± 3.64 & 31.16 ± 3.46 & 29.92 ± 3.45 & 26.58 ± 3.84 & 21.03 ± 1.63 \\ \hline
\end{tabular}
\end{adjustbox}
\vspace{2mm}
\end{table*}

\begin{table*}
\centering
\caption{Impact of data quantity $D_E$ on model accuracy in $\%$.}\label{table-quantity}
\begin{adjustbox}{max width=\textwidth}

\begin{tabular}{|l|c|c|c|c|} \hline
\multicolumn{5}{|c|}{CIFAR10} \\ \hline
& $D_E=3k$ & $D_E=5k$ & $D_E=8k$ & $D_E=10k$ \\ \hline
$I$ local & 38.47 ± 9.91 & 38.63 ± 8.10 & 39.80 ± 8.08 & 40.56 ± 8.25 \\ \hline
$E$ local & 48.07 ± 8.99 & 49.27 ± 2.84 & 51.84 ± 0.66 & 53.23 ± 0.77 \\ \hline
FedAvg & 59.69 ± 9.55 & 60.66 ± 1.62 & 62.12 ± 3.63 & 62.88 ± 5.50 \\ \hline
\end{tabular}

\vspace{2em} 

\begin{tabular}{|l|c|c|c|c|} \hline
\multicolumn{5}{|c|}{CIFAR100} \\ \hline
& $D_E=3k$ & $D_E=5k$ & $D_E=8k$ & $D_E=10k$ \\ \hline
$I$ local & 17.44 ± 2.35 & 17.20 ± 1.61 & 18.64 ± 3.03 & 19.85 ± 4.60 \\ \hline
$E$ local & 16.66 ± 2.17 & 18.30 ± 1.16 & 21.50 ± 3.84 & 22.92 ± 4.87 \\ \hline
FedAvg & 25.40 ± 2.95 & 26.14 ± 0.41 & 29.10 ± 1.15 & 32.08 ± 1.45 \\ \hline
\end{tabular}

\vspace{2em} 

\begin{tabular}{|l|c|c|c|c|} \hline
\multicolumn{5}{|c|}{HAM10000} \\ \hline
& $D_E=2k$ & $D_E=3k$ & $D_E=4k$ & $D_E=5k$ \\ \hline
$I$ local & 76.37 ± 1.65 & 76.28 ± 0.52 & 75.61 ± 0.87 & 75.68 ± 1.24 \\ \hline
$E$ local & 73.67 ± 1.87 & 76.88 ± 1.44 & 77.99 ± 1.24 & 78.11 ± 1.29 \\ \hline
FedAvg & 79.90 ± 0.91 & 80.50 ± 0.72 & 80.89 ± 0.99 & 81.09 ± 1.48 \\ \hline
\end{tabular}

\end{adjustbox}
\vspace{2mm}
\end{table*}
\begin{algorithm}[t]
	\caption{Optimization of Period 1 Pricing}  
	\label{alg:B}  
	\begin{algorithmic}[1] 
		\STATE \textbf{initialize}  $ o_I, o_E, w_q, w_p, w_\phi$. Let $p_{I,1,a}=p_{I,1,b}=p_{I,1,c}=0$.
  
           \vspace{1mm}
            \WHILE{not convergent}
                \STATE Compute $\text{NE}_p\left(p_{I,1,a}, 0, (w_qq_{I,1}-1)/{w_p}\right)$ and $\text{NE}_{r}\left(p_{I,1,a},0, (w_qq_{I,1}-1)/{w_p}\right)$ 
                \STATE Update $p_{I,1,a}$ using gradient ascent
            \ENDWHILE
            \STATE Return converged solution as $p^*_{I,1,a}$

            \vspace{1mm}
            \WHILE{not convergent}
                \STATE Compute $\text{NE}_p\left(p_{I,1,b}, (w_qq_{I,1}-1)/{w_p}, w_qq_{I,1}/w_p\right)$  and $\text{NE}_{r}\left( p_{I,1,b},(w_qq_{I,1}-1)/{w_p}, w_qq_{I,1}/w_p\right)$ 
                \STATE Update $p_{I,1,b}$ using gradient ascent
            \ENDWHILE
            \STATE Return converged solution as $p^*_{I,1,b}$

                        \vspace{1mm}
            \WHILE{not convergent}
                \STATE Compute $\text{NE}_p\left(p_{I,1,c},  w_qq_{I,1}/w_p, \infty\right)$  and $\text{NE}_{r}\left( p_{I,1,c}, w_qq_{I,1}/w_p, \infty\right)$ 
                \STATE Update $p_{I,1,c}$ using gradient ascent
            \ENDWHILE
            \STATE Return converged solution as $p^*_{I,1,c}$
            
             \vspace{1mm}
            \STATE $p^*_{I,1}\leftarrow \arg\max\left\{W_I(p^*_{I,1,a}), W_I(p^*_{I,1,b}), W_I(p^*_{I,1,c})\right\}$.
	\end{algorithmic}  
\end{algorithm}

\section{Numerical Results}\label{numerical}
We conduct numerical experiments to gain more useful insights.  In Sec.~\ref{sec: setup}, we discuss the simulation setup. In Sec.~\ref{sec: training-results}, we train FL models on three different datasets and report the test accuracy. In Sec.~\ref{sec: NE-result}, we use the accuracy results to compute the solutions to the two-period game. 

\subsection{Simulation Setup}\label{sec: setup}
We train ResNet-18  on CIFAR-10 and HAM10000, as well as train ResNet-50 on CIFAR-100  using FedAvg \cite{mcmahan2017communication}.\footnote{Our model is compatible with any FL algorithms, and this paper focuses on FedAvg, which till today, remains one of the state-of-the-art FL algorithms.} CIFAR-10 (100) is a balanced dataset on $10 (100)$ classes with $50k$ training and $10k$ test data \cite{krizhevsky2009learning}. HAM10000 is an imbalanced medical dataset on 7 classes with $10,015$ images from four domains, where each domain represents an institution who collected dermoscopic (skin disease) images \cite{tschandl2018ham10000}.  We consider that the two companies have heterogeneous data, which are sampled using the Dirichlet distribution with a controlling parameter $\beta>0$, where a smaller $\beta$ implies higher data heterogeneity \cite{hsu2019measuring}. We study two scenarios:
\begin{itemize}
\item \textit{Impact of data heterogeneity}: On CIFAR-10 and CIFAR-100, we use $\beta\in\{\infty, 1, 0.5, 0.1, 0.01\}$. For each $\beta$, we sample $5k$ data for company $I$ and $5k$ data for company $E$. We do not explore the impact of data heterogeneity on HAM10000 as the dataset itself is highly imbalanced. 
\item \textit{Impact of data quantity}: On CIFAR-10 and CIFAR-100, we fix $\beta=0.1$, assign company $I$ $5k$ data, and change company $E$'s data volume $D_E\in\{3k,5k,8k, 10k\}$. On HAM10000, we randomly assign company $I$ $2k$ data and company $E$ data of size $ D_E\in \{ 2k, 3k, 4k, 5k\}$.
\end{itemize}
The key hyper-parameters are summarized as follows. We use $100$ rounds for FL training, where each round consists of $5$ local epochs. We use SGD as the optimizer, and choose learning rate  $l_r=0.001$ and batch size $B=64$. We further report the results when each company only uses its own data to train a local model.   Each experiment is repeated over 3 runs using different random seeds. 

\subsection{Training Results}\label{sec: training-results}
 Table \ref{table-beta} reports the training results under different levels of data heterogeneity,  where \textit{``$I$ (E) local''} means that company $I$ ($E$) trains a model using its local data without FL. From this table, we make two observations. 
 
 First, on both datasets, FedAvg consistently outperforms local training models across all values of $\beta$, including the highly heterogeneous case $\beta=0.01$. This suggests the benefit of FL collaboration in improving local model performances. Second, as $\beta$ decreases, there is a general trend of performance decline for FedAvg. This is due to the notorious client drift issue in FL, which has been observed in many distributed learning literature (e.g., \cite{woodworth2020minibatch,li2020federated}). Despite that data heterogeneity hurts FL performance, we will show a counter-intuitive result in Sec.~\ref{sec: NE-result} that data heterogeneity promotes price competition.

 Table \ref{table-quantity} reports the accuracy results under different data quantities. We observe that FedAvg again outperforms local training across different data quantities on the three datasets. Also, increasing data volume $D_E$ consistently improves model accuracy across all datasets, underscoring the importance of data quantity in model training.

\subsection{Equilibrium Results}\label{sec: NE-result}
Now, we use Tables \ref{table-beta}-\ref{table-quantity}  to calculate the equilibrium of the two period game. We consider that companies' quality of service is an increasing function in the model accuracy. In particular, we use $q_{i,t}(A)=A$, where $A$ is the mean accuracy value from the tables. We further consider that users' type (preference) $\phi_n$ is heterogeneous, which follows a uniform distribution on support $[0,1]$.\footnote{We have also tested Gaussian distributions and results carry over.} We choose coefficients $w_p=w_{\phi}=w_p=1$ and normalize the service costs $c_I$ and $c_E$ to $0$. We use Proposition \ref{collaboration-NE} to calculate the equilibrium FL collaboration, in which
we use the best response algorithm \cite{osborne1994course} to compute the equilibrium of Game \ref{price-game}. We further use Algorithm \ref{alg:B} to compute the optimal period 1 pricing of company $I$.

\begin{table}[t]
  \centering
    \caption{FL collaboration under different $\beta$.}
  \begin{tabular}{|c|c|c|c|c|c|}
    \hline
     & $\beta=\infty$ & $\beta=1$ & $\beta=0.5$ & $\beta=0.1$ & $\beta=0.01$ \\
    \hline
    CIFAR-10 & $\checkmark$ & $\checkmark$ & $\checkmark$ & $\checkmark$ & $\checkmark$ \\
    \hline
    CIFAR-100 & $\checkmark$ & $\checkmark$ & $\checkmark$ & $\checkmark$ & $\checkmark$ \\
    \hline
  \end{tabular}
  \label{table-FL-beta}
\end{table}

\begin{table}[t]
  \centering
    \caption{FL collaboration under different $D_E$.}
    \begin{adjustbox}{max width=\textwidth}
  \begin{tabular}{|c|c|c|c|c|}
    \hline
     & $D_E=3k$ & $D_E=5k$ & $D_E=8k$ & $D_E=10k$ \\
    \hline
    CIFAR-10 & $\checkmark$ & $\checkmark$ & $\checkmark$ & $\checkmark$  \\
    \hline
    CIFAR-100 & $\checkmark$ & $\checkmark$ & $\checkmark$ & $\checkmark$ \\
    \hline
     & $D_E=2k$ & $D_E=3k$ & $D_E=4k$ & $D_E=5k$ \\
     \hline
    HAM10000 & $\times$ & $\times$ & $\times$ & $\times$ \\
    \hline
  \end{tabular}
\end{adjustbox}
  \label{table-FL-quantity}
\end{table}

\begin{figure*}
	\centering
        \subfloat[Impact of $\beta$ on CIFAR-10.]{\includegraphics[width=1.65in]{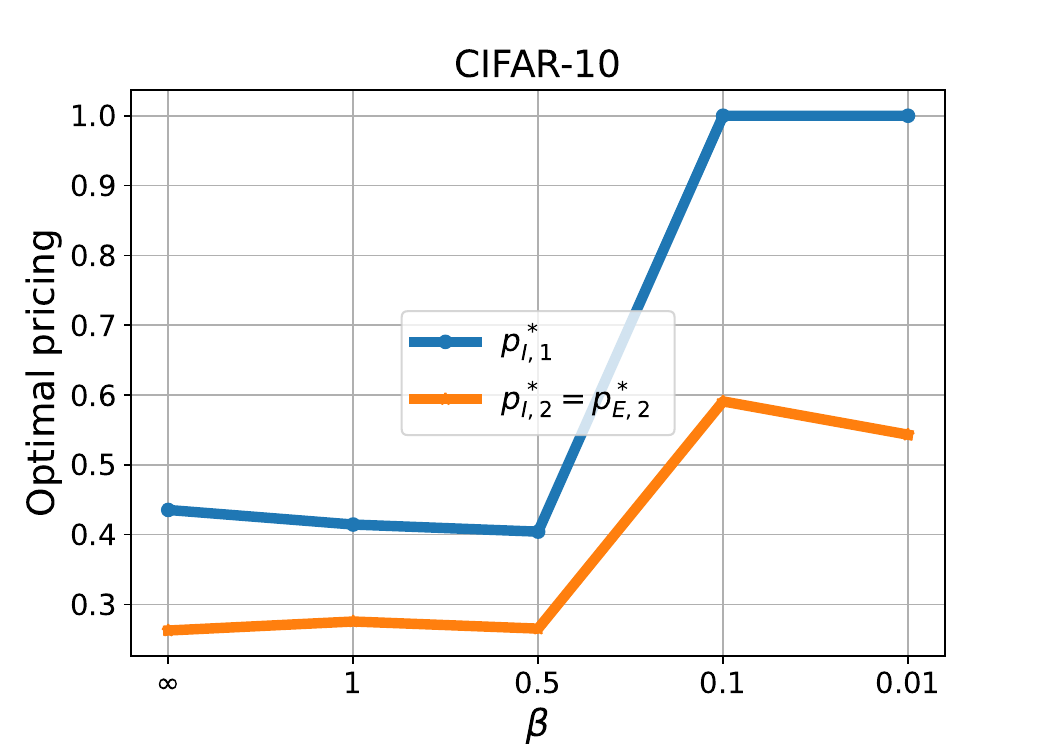}}
	\hfil
  \subfloat[Impact of $\beta$ on CIFAR-100.]{\includegraphics[width=1.65in]{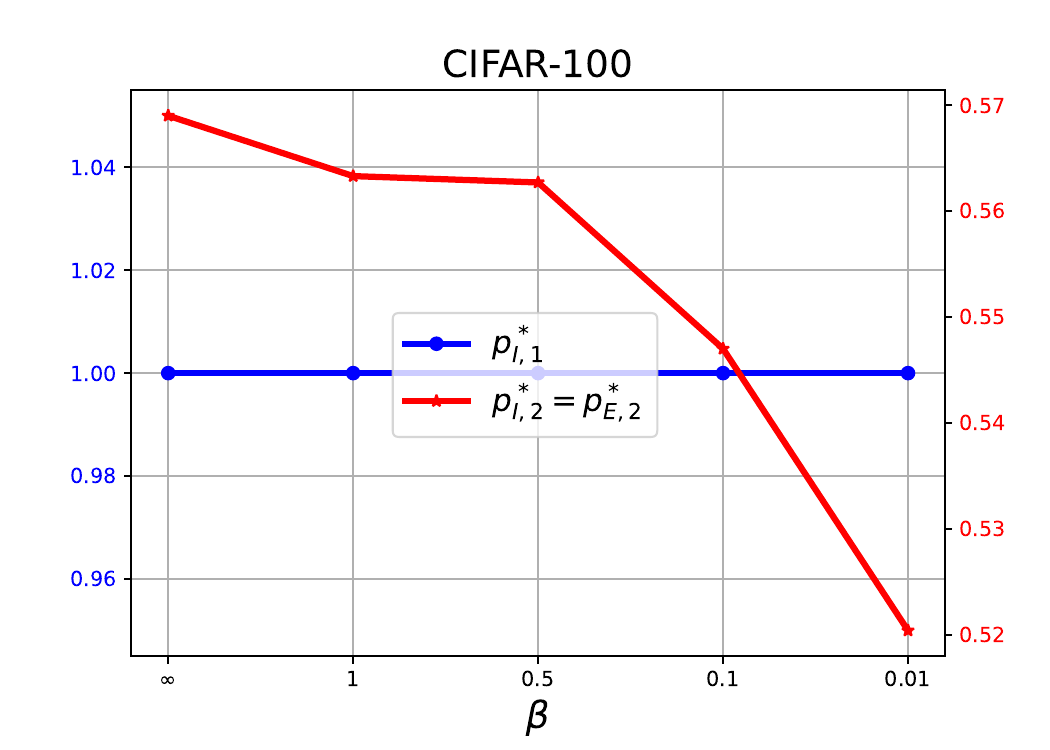}}
	\hfil
	\subfloat[Impact of $D_E$ on CIFAR-100.]{\includegraphics[width=1.65in]{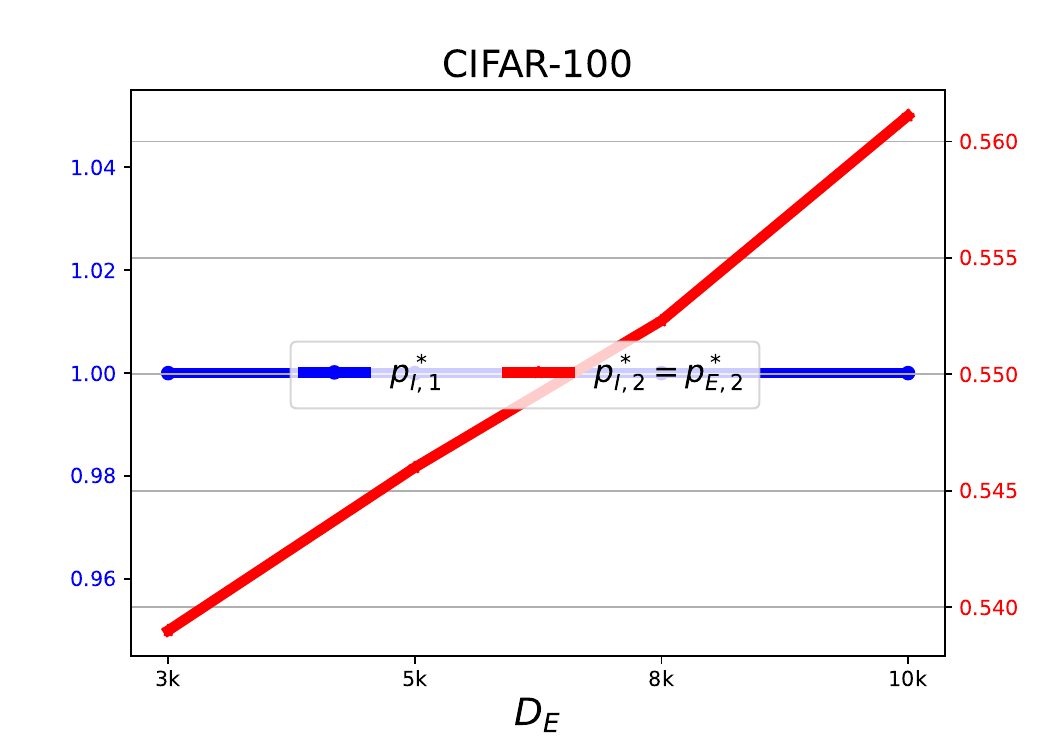}}
	\hfil
	\subfloat[Impact of $D_E$ on HAM10000.]{\includegraphics[width=1.65in]{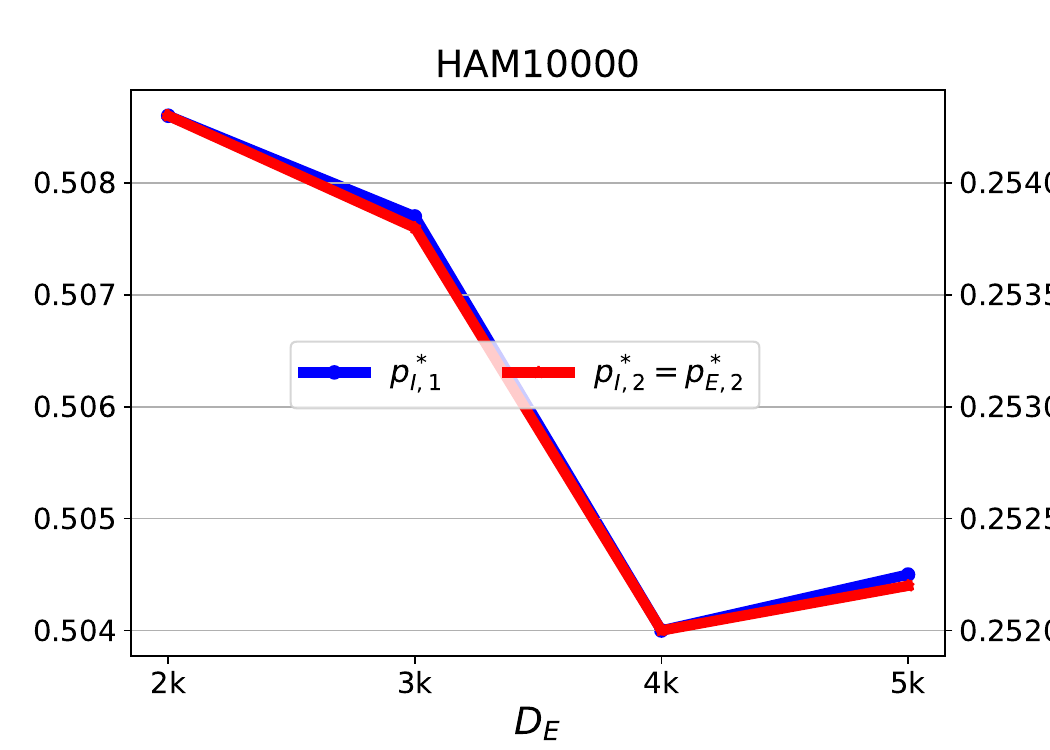}}
	\hfil
\vspace{2mm}
	\caption{Optimal two-period pricing on different datasets.}
	\label{fig-two-period-pricing}
 \vspace{3mm}
\end{figure*}


\textbf{Optimal FL collaboration}. We first investigate the equilibrium FL collaboration and report the results in Tables \ref{table-FL-beta}-\ref{table-FL-quantity}, where $\checkmark$ means collaboration and $\times$ means no collaboration. We observe that while FL enhances model performance across the three datasets (see Table \ref{table-beta}), intriguingly, collaboration is observed only on CIFAR datasets but not on HAM10000. This may seem counter-intuitive at first glance because the improved performance through FL suggests a natural incentive to collaborative behavior. However, it is crucial to note that improved performance alone does not necessitate collaboration in a competitive environment. Recall from Corollary \ref{cor-price-quality-relation} that a company's price increases in its own model performance but at the same time decreases in its competitor's model performance. Hence, collaboration is a beneficial strategy only when the performance gains are substantial enough to outweigh the loss from price competition. For instance, CIFAR datasets exhibit significant performance enhancements (e.g., $\beta=0.1$ on CIFAR-10 in Table \ref{table-beta}), which motivates collaboration, while the less significant improvements in HAM10000 (e.g., $D_E=3k$ in Table \ref{table-quantity}) are insufficient to motivate the same collaborative effort. We summarize the above observation below. 
\begin{Obs}
FL collaboration is the NE only when the model performance improvement is substantial.
\end{Obs}

\textbf{Optimal Pricing}. We now investigate how the optimal pricing changes with data heterogeneity $\beta$ and data quantity $D_E$. The results are reported in Fig.~\ref{fig-two-period-pricing}. Note that in period 2 the two companies' optimal pricing are the same due to 1) both companies having the same (global) model, and 2) users' preference $\phi_n$ follows a uniform distribution, which is symmetric for both companies. 

From Fig.~\ref{fig-two-period-pricing}(a)-(b), it is a bit surprising to observe that both companies optimal pricing can  increase in data heterogeneity (i.e., a smaller $\beta$).  While the expectation is that higher heterogeneity leads to worse performance and hence users are willingly to pay, leading to smaller prices. However, this is not always the case. The reason is that when data is highly heterogeneous, even if the model performance declines, the companies still obtain a big performance improvement from FL compared to local learning (e.g., see $\beta=0.1$ in Table \ref{table-beta}). As a result, company $I$ sets a  high price in period 1 so that no users buy low-quality service in this period. For example, in Fig.~\ref{fig-two-period-pricing}(a) at $\beta=0.1$, $I$ sets $p^*_{I,1}=1$ and based on Lemma \ref{user-period-1}, none of the users will purchase from $I$. Instead, both companies attract users to buy high-quality services at higher prices in period 2, which results in higher profits. This observation has an interesting implication, i.e., data heterogeneity does not always prevent but can promote price competition. In particular, the incumbent company $I$ does not harvest most users in the starting period but prefers to compete with the entrant company using better model-based services. We summarize the above observations as follows.
\begin{Obs}
Data heterogeneity can promote price competition between incumbent and entrant companies. 
\end{Obs}

From Fig.~\ref{fig-two-period-pricing}(c)-(d), we observe different trends of pricing as a function of $D_E$ on the two datasets. The reason is that the optimal FL collaboration decisions are different (refer to Table \ref{table-FL-quantity}). On CIFAR-100, the companies collaborate via FL and the performance improvement tends to be larger as $D_E$ increases. Hence, company $I$ uses a very large price in period 1 so that no users buy in period 1, and instead users will buy higher-quality services in period 2 at a larger price.  On HAM10000, however, the companies do not collaborate in FL. As a result, company $I$ is inclined to set a smaller price in period 1 so that most users subscribe to $I$ in the first period. Then in period 2, the companies participate in more fierce price competition due to a smaller user pool, leading to a smaller price.

\section{Discussions}\label{Discussions}
We have provided a first study on FL coopetition using a dynamic two-period model with two companies. The analysis is highly non-trivial with this stylized model involving a non-concave problem. Next, we discuss how to extend our model with more than two companies across multiple time periods. 

\textbf{Extension to multiple companies}. Our model can be extended to scenarios with two or more companies. In particular, consider $N\ge 3$ companies where $N-1$ of them coexist in period 1, and an entrant enters in period 2. This can model real-world market conditions where multiple companies often coexist and new entrants periodically disrupt existing equilibriums. In this case, we can similarly analyze the user behavior by comparing the payoffs obtained when purchasing from different companies. For companies' FL collaboration, we can use the coalitional game theory to find stable FL collaboration structures, e.g., \cite{donahue2021optimality,tan2024fedcompetitors}. For companies' price competitions, prior studies on oligopoly competition \cite{tirole1988theory} can offer theoretical and algorithmic solutions. However, how to optimally set prices in period 1 (which affects period 2 outcomes) remains a challenging but interesting problem to explore in future work. 

\textbf{Extension to multiple time periods}. It is possible to extend our model to where there are $t\ge3$ periods, and in each period a new entrant company enters the market. For user decisions, it suffices to restrict attention in each period and compute the highest payoff achieved among existing companies. However, we note that solving companies' FL collaboration and price competition over $t\ge 3$ periods in closed-form is generally analytically intractable. One particular challenge is to estimate future entrants and their properties (e.g., data distributions). One possible remedy is to resort to reinforcement learning for sequential decision making, and in particular, multi-armed bandit \cite{slivkins2019introduction,huang2023online}. More specifically, one could model the \textit{agents} as companies, the \textit{arms} as FL collaboration and pricing, and the \textit{reward signals} as profits achieved from users, which takes into account competitions and model improvement from FL training.

\section{Conclusion}\label{Conclusions}
This work studied the under-explored problem of FL coopetition (FL collaboration and market competition) using a dynamic two-period model. One challenge pertains to multi-dimensional heterogeneity in terms of companies' data distributions and quantities, as well as users' service preferences. Another challenges is associated with solving a non-concave optimization problem. We decomposed the problem into multiple concave sub-problems and managed to characterize the solutions to the two-period game. Theoretical analysis and numerical experiments on three datasets have lead to two implications. First, FL brings both model performance gain and competition loss, and collaboration occurs only when the performance gain outweighs the loss.  Second, data heterogeneity can incentivize the incumbent to limit market penetration in period 1 and promote price competition in period 2. 

For future work, it is interesting to study how data quality (e.g., label correctness and label sparsity) affects FL coopetition and further develop robust algorithms to enhance performance. It is also interesting to incorporate privacy enhancing techniques, which can be of particular interest to companies, into the coopetition analysis. 



\onecolumn 

\section{Appendices}
We organize the supplementary materials as follows. 
\begin{itemize}
	\item In Appendix \ref{proof-lemma-1}, we prove Lemma 1.
	\item In Appendix \ref{proof-lemma-2}, we prove Lemma 2.
	\item In Appendix \ref{proof-proposition-1}, we prove Proposition 1.
	\item In Appendix \ref{proof-corollary-1}, we prove Corollary 1.
	\item In Appendix \ref{proof-proposition-2}, we prove Proposition 2. 
	\item In Appendix \ref{proof-theorem-1}, we prove Theorem 1. 
	\item In Appendix \ref{proof-theorem-2}, we prove Theorem 2.
	\item In Appendix \ref{best-response}, we present a best response algorithm to calculate the NE of Game 1.	
\end{itemize}


\section{Proof of Lemma 1}\label{proof-lemma-1}
\begin{proof}
	Consider the payoff function for user \(n\) in period \(t=1\) given by:
	\[
	u_{n,t} (d_{n,t}; \boldsymbol{r}, \boldsymbol{p}) = 
	\begin{cases} 
	0, & \text{if } d_{n,t} = \emptyset,\\
	w_q q_{I, t}(\boldsymbol{r}) - w_\phi \phi_n - w_p p_{I,t}, & \text{if } d_{n,t} = I.
	\end{cases}
	\]
	
	At \(t=1\), the user can either choose to purchase from \(I\) or not purchase. 
	
	For purchasing from \(I\), the utility is:
	\[
	u_{n,1}(I; \boldsymbol{r}, \boldsymbol{p}) = w_q q_{I,1}(\boldsymbol{r}) - w_\phi \phi_n - w_p p_{I,1}.
	\]
	
	For the user to prefer purchasing over not purchasing, we require:
	\[
	w_q q_{I,1}(\boldsymbol{r}) - w_\phi \phi_n - w_p p_{I,1} \ge 0.
	\]
	
	Thus, the condition for user \(n\) to choose \(I\) over \(\emptyset\) at \(t=1\) is (when we normalize $w_\phi=1$):
	\[
	\phi_n \leq w_q q_{I,1} - w_p p_{I,1}.
	\]
	
	This, we finish the proof for Lemma 1. 
\end{proof}

\section{Proof of Lemma 2}\label{proof-lemma-2}
\begin{proof}
	
	Given the assumption \(w_q q_{I,2}(\boldsymbol{r}) - w_p p_{I,2} \geq w_q q_{E,2}(\boldsymbol{r}) - w_p p_{E,2}\), purchasing from $I$ is more efficient from a ``cost-benefit'' perspective. Moreover, a user will buy in period 2 if it does not buy in period 1, i.e., $d_{n,1}=\emptyset.$
	
	For user \(n\) to choose \(I\) in period 2, the payoff should be larger than that from not buying or buying from $E$. That is
	\begin{displaymath}
	w_q q_{I, t}(\boldsymbol{r}) - w_\phi \phi_n - w_p p_{I,t}\ge \max\left\{0, w_q q_{E, t}(\boldsymbol{r}) - w_\phi (1-\phi_n) - w_p p_{E,t}\right\}. 
	\end{displaymath}
	
	Rearranging this, \(\phi_n\) should be within:
	\[
	\phi_n \leq \max\left(\min\left(w_q q_{I,2}(\boldsymbol{r}) - w_p p_{I,2}, \frac{1 + w_q (q_{I,2}(\boldsymbol{r}) - q_{E,2}(\boldsymbol{r})) - w_p (p_{I,2} - p_{E,2})}{2}\right), 1\right).
	\]
	
	To choose \(E\), \(\phi_n\) should be high, reflecting a strong preference misalignment favoring \(E\) despite its lower net utility. Similarly, we obtain:
	\[
	\phi_n \geq \min\left(\max\left(1 - w_q q_{E,2}(\boldsymbol{r}) + w_p p_{E,2}, \frac{1 + w_q (q_{I,2}(\boldsymbol{r}) - q_{E,2}(\boldsymbol{r})) - w_p (p_{I,2} - p_{E,2})}{2}\right), 0\right).
	\]
	
	If neither condition for \(I\) nor \(E\) is satisfied, user \(n\) will not purchase, thus choosing \(\emptyset\).
	
	Thus, we finish the proof.
\end{proof}

\section{Proof of Proposition 1}\label{proof-proposition-1}
\begin{proof}
	We need a useful mathematical result to prove Lemma 2. We state it as a lemma below.
	\begin{Lem}\label{existence}
		There exists a Nash equilibrium  if 
		\begin{enumerate}
			\item The company set is finite.
			\item The strategy space is closed, bounded, and convex.
			\item The company profit functions are continuous and quasi-concave in their strategies.
		\end{enumerate}
	\end{Lem}
	One can easily check that Game 1 has a finite company set $\mathcal{N}=\{I,E\}$. Note that even if each company's strategy $p_{i,t}\ge 0$ is unbounded, we can narrow down the strategy space to $[0, w_qq_{i,t}]$, as any $w_qq_{i,t}$ leads to a zero profit. In this case,  the reduced strategy space for each company is closed, bounded, and convex. It remains to show that each company's profit function is quasi-concave in $p_{i,t}$.  
	We begin by deriving the decision threshold $\phi^*$ for a user to choose organization I over E. Given the user utility functions and assuming:
	\[
	w_q q_{I,2} - w_\phi \phi - w_p p_{I,2} \geq w_q q_{E,2} - w_\phi (1 - \phi) - w_p p_{E,2}
	\]
	Rearranging the terms gives:
	\[
	w_\phi (1 - 2\phi) \leq w_q (q_{I,2} - q_{E,2}) - w_p (p_{I,2} - p_{E,2})
	\]
	Solving for $\phi$, we find an indifference point:
	\[
	\phi^* = \frac{1}{2} \left(1 + \frac{w_q (q_{I,2} - q_{E,2}) - w_p (p_{I,2} - p_{E,2})}{w_\phi}\right)
	\]
	
	The profit function $W_{I,2}$ can be written as:
	\[
	W_{I,2} = (p_{I,2} - c_I) H(\phi^*)
	\]
	Taking the first derivative with respect to $p_{I,2}$, we get:
	\[
	\frac{\partial W_{I,2}}{\partial p_{I,2}} = H(\phi^*) + (p_{I,2} - c_I) h(\phi^*) \frac{\partial \phi^*}{\partial p_{I,2}}.
	\]

	Now, taking the second derivative, we find:
	\[
	\frac{\partial^2 W_{I,2}}{\partial p_{I,2}^2} = (p_{I,2} - c_I) \left( h'(\phi^*) \left(\frac{\partial \phi^*}{\partial p_{I,2}}\right)^2 + h(\phi^*) \frac{\partial^2 \phi^*}{\partial p_{I,2}^2} \right)
	\]
	Given that $\frac{\partial \phi^*}{\partial p_{I,2}}\le0$, $\frac{\partial^2 \phi^*}{\partial p^2_{I,2}}=0$, $h'(\phi^*) \leq 0$, we can see $\frac{\partial^2 W_{I,2}}{\partial p_{I,2}^2}\le 0$. Thus, $W_{I,2}$ is concave in $p_{I,2}$. 
	
	A similar process can be applied for $W_{E,2}$.
	Assuming the utility functions dictate user decisions based on:
	\[
	w_q q_{I,2} - w_\phi \phi - w_p p_{I,2} \leq w_q q_{E,2} - w_\phi (1-\phi) - w_p p_{E,2},
	\]
	we can rearrange this inequality to find the cutoff value \(\phi^*\) where users switch preference from I to E:
	\[
	w_\phi (1 - 2\phi) \geq w_q (q_{E,2} - q_{I,2}) - w_p (p_{E,2} - p_{I,2}),
	\]
	\[
	\phi^* = \frac{1}{2} \left(1 + \frac{w_q (q_{E,2} - q_{I,2}) - w_p (p_{E,2} - p_{I,2})}{w_\phi}\right).
	\]
	
	The profit function \( W_{E,2} \) can then be expressed as:
	\[
	W_{E,2} = (p_{E,2} - c_E) \int_{\phi^*}^{1} h(\phi) d\phi = (p_{E,2} - c_E) (1 - H(\phi^*)).
	\]
	
	First, compute the first derivative:
	\[
	\frac{\partial W_{E,2}}{\partial p_{E,2}} = 1 - H(\phi^*) + (p_{E,2} - c_E) h(\phi^*) \frac{\partial \phi^*}{\partial p_{E,2}}.
	\]
	
	indicating that as \( p_{E,2} \) increases, \(\phi^*\) increases, reducing the interval over which \(E\) is chosen.
	
	Now, compute the second derivative:
	\[
	\frac{\partial^2 W_{E,2}}{\partial p_{E,2}^2} = (p_{E,2} - c_E) \left( h'(\phi^*) \left(\frac{\partial \phi^*}{\partial p_{E,2}}\right)^2 + h(\phi^*) \frac{\partial^2 \phi^*}{\partial p_{E,2}^2} \right),
	\]
	Note that \( h'(\phi^*) \leq 0 \) and \( h(\phi^*) \geq 0 \), the terms involving derivatives of \( h \) indicate a negative value for the second derivative when the interval \( [\phi^*, 1] \) decreases as \( p_{E,2} \) increases. Thus, \( W_{E,2} \) is concave in \( p_{E,2} \). 
	
	Hence we finish the proof for Proposition 1.
\end{proof}

\section{Proof of Corollary 1}\label{proof-corollary-1}
\begin{proof}
	\textit{Part 1: Increasing Function of Its Own Quality \( q_{i,t} \)}
	
	Assume for contradiction that as \( q_{i,t} \) increases, \( p_{i,t} \) decreases. If \( q_{i,t} \) increases, this  leads to increased user valuation for company \( i \)'s product, which increases the demand at any given price level. However, if \( p_{i,t} \) decreases while \( q_{i,t} \) increases, the company misses the opportunity to maximize its profit by leveraging the higher perceived value of a higher quality product. This contradicts profit maximization, thus our assumption must be false. Hence, \( p_{i,t} \) must be an increasing function of \( q_{i,t} \).
	
	\textit{Part 2: Decreasing Function of Competitor's Quality \( q_{j,t} \)}
	
	Assume for contradiction that as \( q_{j,t} \) increases, \( p_{i,t} \) increases. An increase in \( q_{j,t} \) enhances the competitor \( j \)'s product attractiveness, likely diverting users away from \( i \)'s product, especially if \( p_{i,t} \) also increases. This would lead to a decrease in the demand for \( i \)'s product and thus a reduction in its profits. To counterbalance the increase in \( q_{j,t} \) and maintain competitiveness, rational behavior would dictate that \( i \) should decrease \( p_{i,t} \) to retain market share and profit margins. Therefore, our assumption leads to a contradiction with the profit-maximizing behavior expected in competitive markets. Hence, \( p_{i,t} \) must decrease as \( q_{j,t} \) increases.
	
	Thus, we conclude that the equilibrium pricing \( p_{i,t} \) is an increasing function of its own quality \( q_{i,t} \) and a decreasing function of the competitor's quality \( q_{j,t} \).
\end{proof}

\section{Proof of Proposition 2}\label{proof-proposition-2}
\begin{proof}
	The proof of Proposition 2 is immediate given that in a two-company case, if any one of the company does not participate in FL, then FL ceases to occur. Hence, based on Definition 2, we know that the profile $(1, 1)$ is the NE of Game 2 if and only if $\forall i\in \left\{I,E\right\}$, $W_{i,2}( (1,1), \boldsymbol{p}_2^*((1,1), p_{I,1})) \ge W_{i,2}((0, 0), \boldsymbol{p}_2^*((0,0), p_{I,1})).$
\end{proof}

\section{Proof of Theorem 1}\label{proof-theorem-1}
\begin{proof}
	We prove Theorem 1 by studying a concrete scenario using a uniform distribution of users' preferences. It consists of three steps. 
	\begin{enumerate}
		\item  Uniform distribution satisfy the assumptions.
		\item  Use a particular value of $p_{I,1}$ and derive the users' purchasing in period 1. 
		\item  Construct accuracy scenarios that are consistent with Theorem 1.
	\end{enumerate}
	
	1. For the uniform distribution, we know that \( h(\theta) > 0 \) and is continuous:
	\begin{itemize}
		\item \( h(\phi_n) = 1 >0\) 
		\item The function \( h(\phi_n) \) is constant over its domain, hence it is continuous.
	\end{itemize}
	Further, to show whether \( \frac{h(\phi_n)}{1 - H(\phi_n)} \) is increasing in \( \phi_n \), we calculate
	\begin{displaymath}
	\frac{h(\phi_n)}{1 - H(\phi_n)} = \frac{1}{1 - \phi_n},
	\end{displaymath}
	which increases in $\phi_n$. Hence, the uniform distribution satisfies Assumption 1. 
	
	2. We consider that $p_{I,1}\ge w_qq_{I,1}/w_p$ so that no users will buy in period 1. 
	
	3. Now we construct a scenario where local learning leads to $q_{I,2}(0,0)=0.72, q_{E,2}(0,0)=0.73$ , while FL training leads to $q_{I,2}(1,1)=0.75, q_{E,2}(1,1)=0.75$. This implies that FL improves both companies' qualities. Now, consider $w_q=w_p=w_\phi=1, c_I=C_E=0$, and report the equilibrium profit (calculated using best response in Section 8 which according to \cite{osborne1987equilibrium}  converges to NE.)

	\begin{center}
		\begin{tabular}{|c|c|c|}
			\hline
			- & $r_I=1$ & $r_E=0$ \\
			\hline
			$r_I=1$ & $W_{I,2}=0.061, W_{E,2}=0.0675$ & - \\
			\hline
			$r_E=0$& - & $W_{I,2}=0.0624, W_{E,2}=0.0624$ \\
			\hline
		\end{tabular}
	\end{center}
	From this table, we see that even if participating in FL leads to higher accuracy for both qualities, the profit of both company is lower, resulting in non-participation as equilibrium.  
	
	Hence, we complete the proof for Theorem 1. 
\end{proof}

\section{Proof of Theorem 2}\label{proof-theorem-2}
We aim to prove the concavity in three sub-problems (corresponding to different regions).
\begin{enumerate}
	\item Case a: $p_{I,1}\in [0,(w_qq_{I,1}-1)/{w_p}]$
	\item Case b: $p_{I,1}\in [(w_qq_{I,1}-1)/{w_p}, w_qq_{I,1}/{w_p}]$
	\item Case c: $p_{I,1}\in [w_qq_{I,1}/{w_p}, \infty]$
\end{enumerate}

1. We first prove Case a. When $p_{I,1}\in [0,(w_qq_{I,1}-1)/{w_p}]$, it is easy to show that all users will buy from company $I$. Hence, 
\begin{displaymath}
W_I=W_{I,1}(p_{I,1}) = \int_{0}^{1} (p_{I,1}-c_{I}) \cdot \mathbbm{1}_{d_{n,1}=I}(p_{I,1}) \cdot h(\phi) d\phi=p_{I,1}-c_I,
\end{displaymath}
which is clearly concave in $p_{I,1}$.

2. Now we prove case b. e analyze the concavity of \( W_I \) by examining its components \( W_{I,1}(p_{I,1}) \) and \( W_{I,2}(\boldsymbol{r}, \boldsymbol{p}) \).

Analysis of \( W_{I,1}(p_{I,1}) \):
\( W_{I,1}(p_{I,1}) \) is given by:
\[
W_{I,1}(p_{I,1}) = \int_{0}^{1} (p_{I,1} - c_{I}) \cdot \mathbbm{1}_{d_{n,1}=I}(p_{I,1}) \cdot h(\phi) d\phi,
\]
where \( \mathbbm{1}_{d_{n,1}=I}(p_{I,1}) = 1 \) if \( w_p p_{I,1} \leq w_q q_{I,1} \), and 0 otherwise. This corresponds to a concave function.

Analysis of \( W_{I,2}(\boldsymbol{r}, \boldsymbol{p}) \):
\( W_{I,2}(\boldsymbol{r}, \boldsymbol{p}) \) depends on \( p_{I,1} \) linearly as $p_{I,1}$ only affects user range in period 2. 

Combining the Functions:
Since both components are concave and \( W_{I,2} \) does not depend on \( p_{I,1} \), the sum \( W_I(\boldsymbol{r}, \boldsymbol{p}) = W_{I,1}(p_{I,1}) + W_{I,2}(\boldsymbol{r}, \boldsymbol{p}) \) retains concavity with respect to \( p_{I,1} \), as the addition of a constant (or concave independent term) to a concave function does not affect its concavity. Hence, \( W_I \) is concave in \( p_{I,1} \).

3. Now we prove case c. When $p_{I,1}\ge w_{q}q_{I,1/w_p}$, no one buys in period 1, i.e., $W_{I,1}=0$. $W_{I,2}$ trivially does not depend on $p_{I,1}$. Hence, $W_I$ is also concave in this range.

Thus, we finish the proof of Theorem 2.

\section{Best Response Algorithm}\label{best-response}
We provide a best response algorithm below to calculate the equilibrium of Game 1. 

\begin{algorithm}
	\caption{Best Response Algorithm for Price Competition Game}
	\begin{algorithmic}[1]
		\STATE Initialize $p_{I,2}^{(0)}$ and $p_{E,2}^{(0)}$ arbitrarily
		\STATE  Set convergence tolerance $\epsilon > 0$
		\STATE  Set $k \gets 0$
		\REPEAT
		\STATE  $k \gets k + 1$
		\STATE  Compute $p_{I,2}^{(k)}$ as:
		\[
		p_{I,2}^{(k)} = \arg\max_{p_{I,2}} \int_{0}^{1} (p_{I,2} - c_{I}) \cdot \mathbbm{1}_{d_{n,2}=I}(\boldsymbol{r}, \boldsymbol{p}) \cdot h(\phi) d\phi
		\]
		\STATE  Compute $p_{E,2}^{(k)}$ as:
		\[
		p_{E,2}^{(k)} = \arg\max_{p_{E,2}} \int_{0}^{1} (p_{E,2} - c_{E}) \cdot \mathbbm{1}_{d_{n,2}=E}(\boldsymbol{r}, \boldsymbol{p}) \cdot h(\phi) d\phi
		\]
		\STATE  Check for convergence:
		\[
		\text{if } |p_{I,2}^{(k)} - p_{I,2}^{(k-1)}| < \epsilon \text{ and } |p_{E,2}^{(k)} - p_{E,2}^{(k-1)}| < \epsilon \text{ then stop}
		\]
		\UNTIL{Convergence}
		\STATE  \textbf{return} $p_{I,2}^{(k)}, p_{E,2}^{(k)}$
	\end{algorithmic}
\end{algorithm}



\bibliography{mybibfile}

\end{document}